\documentclass[twocolumn,rnote]{aa}
\usepackage{graphics}
\usepackage{graphicx}
\usepackage{amsfonts}
\usepackage{subfig}

\usepackage{color,ulem,soul} 

\include{epsf}
\include{epstopdf}
\DeclareGraphicsExtensions{.eps,.jpg,.pdf}
\newcommand{\bfo}[1]{\mbox{\boldmath $#1$}}
\def\bvarphi{\mbox{\boldmath $\varphi$}}

\begin{document}
\newcommand{\beq}{\begin{equation}}
\newcommand{\eeq}{\end{equation}}
\def\la{\hbox{\raise.35ex\rlap{$<$}\lower.6ex\hbox{$\sim$}\ }}
\def\ga{\hbox{\raise.35ex\rlap{$>$}\lower.6ex\hbox{$\sim$}\ }}
\def\runit{\hat {\bf  r}}
\def\phunit{\hat {\bfo \bvarphi}}
\def\etaunit{\hat {\bfo \eta}}
\def\zunit{\hat {\bf z}}
\def\zetaunit{\hat {\bfo \zeta}}
\def\xiunit{\hat {\bfo \xi}}
\def\beq{\begin{equation}}
\def\eeq{\end{equation}}
\def\beqa{\begin{eqnarray}}
\def\eeqa{\end{eqnarray}}
\def\sub#1{_{_{#1}}}
\def\order#1{{\cal O}\left({#1}\right)}
\newcommand{\sfrac}[2]{\small \mbox{$\frac{#1}{#2}$}}

\renewcommand\emph[1]{\textit{#1}}
\definecolor{dark-red}{rgb}{0.75, 0.00, 0.00}
\definecolor{hlcolor}{rgb}{1.00, 0.90, 0.85}
\sethlcolor{hlcolor}

%
%
\title{{Linear analysis of the vertical shear instability: outstanding issues and improved solutions}}

\author{O. M. Umurhan\inst{1,2}, R. P. Nelson \inst{3} and O. Gressel \inst{4} }

   \offprints{O. M. Umurhan\\ \email{orkan.m.umurhan@nasa.gov}}

   \institute{ Space Sciences Division, NASA Ames Research Center,
     Moffett Field, CA 94035. \and SETI Institute, 189 Bernardo Way,
     Mountain View CA 94043, USA \and Queen Mary University of
     London, School of Physics and Astronomy, 327 Mile End Road
     London, E1 4NS, UK \and Niels Bohr International Academy, The
     Niels Bohr Institute, Blegdamsvej 17, DK-2100, Copenhagen \O,
     Denmark }

   \date{}


 \abstract {The Vertical Shear Instability is one of two known
   mechanisms potentially active in the so-called dead zones of
   protoplanetary accretion disks.  A recent analysis by Barker and
   Latter (2015) indicates that a subset of unstable modes shows
   unbounded growth -- both as resolution is increased and when the
   nominal lid of the atmosphere is extended.  This trend possibly
   indicates a certain level of ill-posedness in previous attempts of
   linear analysis. }
 {This research note examines both the energy content of the
   aforementioned modes and questions the legitimacy of assuming
   separable solutions for a problem whose linear operator is
   fundamentally inseparable. }
 {The reduced equations governing the instability are revisited and
   the generated solutions are examined using both the previously
   assumed separable forms and an improved non-separable solution form
   that is herewith introduced. }
 {Reconsidering the solutions of the reduced equations using the
   separable form shows that, while the low-order body modes have
   converged eigenvalues and eigenfunctions (as both the vertical
   boundaries of the atmosphere are extended and with increased radial
   resolution), it is also confirmed that the corresponding high-order
   body modes and the surface modes do indeed show unbounded growth
   rates.  However, the energy contained in both the higher-order body
   modes and surface modes diminishes precipitously due to the disk's
   Gaussian density profile. Most of the energy of the instability is
   contained in the low-order modes. An inseparable solution
   form is introduced which filters out the inconsequential surface
   modes leaving only body modes (both low- and high-order ones).  The
   analysis predicts a fastest growing mode with a specific radial
   length scale.  The growth rates associated with the fundamental
   corrugation and breathing modes matches the growth and length
   scales observed in previous nonlinear studies of the instability. }
 {Linear stability analysis of the vertical shear instability should
   be done assuming non-separable solutions.  It is also concluded
   that the surface modes are relatively inconsequential because of
   the little energy they contain, and are artifacts of imposing
   specific kinematic vertical boundary conditions in isothermals 
   disk models. }

 \titlerunning{Non-separable operators in linear VSI}

 \keywords{Interstellar and circumstellar matter, protoplanetary disks, instabilities, turbulence, waves, methods: analytical}

 \maketitle

\section{Introduction}

The Vertical Shear Instability (VSI) (Urpin 2003, Urpin \& Brandenburg
1998, Arlt \& Urpin 2004, Nelson et al. 2013, McNally \& Pessah, 2014,
Stoll and Kley 2015), sometimes known as the Goldreich Schubert Fricke
instability (Goldreich \& Schubert 1967, Fricke 1968) is a linear
instability of axisymmetric inertial modes relying on the vertical
shear of the basic near-Keplerian flow state.  This instability may be
active in non-magnetized parts of protoplanetary accretion disks and
is perhaps discernible in their dead zones (Turner et al., 2014).
Nelson et al. (2013, NGU13 hereafter) and Stoll and Kley (2014) have
demonstrated that the instability can generate a modest amount of
turbulence, with effective disk $\alpha$ ranging somewhere between
$4\times 10^{-4}$ up to $10^{-3}$.  In both studies, the basic
background setting is that of a locally isothermal disk with a radial
temperature gradient arising either from some external imposition
(NGU13) or naturally manifesting itself due to the inclusion of
radiative transfer effects (Stoll and Kley, 2014).

A satisfactory linear stability analysis is still lacking for the disk
setting.  While the basic essence of the instability has been sketched
out using a local point analysis (Goldreich \& Schubert 1967, Fricke
1968, Urpin 2003), the way the instability manifests itself in a
global or semi-global disk setting is difficult to assess because the
basic linear stability problem is non-separable even in the simplest
model reduction (NGU13, Barker \& Latter 2015, BL15 hereafter).  NGU13
and BL15 present such a linear stability analysis using a reduced
model set and they show that while the low-order modes that go
unstable are consistent with the time scales of the instability seen
in the numerical experiments, there are some serious shortcomings
associated with the analysis that cast doubt as to whether it
provides an accurate description of the physical manifestation
of the VSI, especially when analyzed in the locally isothermal disk 
setting.

The mode analyses done by NGU13 and BL15 show that if one assumes
radially propagating traveling waves, there are two classes of modes
loosely referred to as \emph{body} modes and \emph{surface} modes.
The surface modes come into existence if one imposes no-flow boundary
conditions like an impenetrable lid at positions above and below the
disk midplane (usually at least a few local disk scale heights or
higher). The body modes are present irrespective of the kinematic
conditions in the vertical so long as the kinetic energies decay away
sufficiently far from the midplane (see below).  BL15 also point out
that the surface modes become more multitudinous as the radial
disturbance wavelengths becomes shorter.  BL15 and NGU13 show that for
a given value of the radial wavenumber there exists a mode with the
fastest growth rate which corresponds, generally, to a surface
mode.

\noindent However, there are three troubling features:
\begin{enumerate}\itemsep6pt
\item BL15 point out that, as the nominal lid of the atmosphere is
  extended to infinity, the fastest growing eigenmodes have growth
  rates that become similarly unbounded as well, growing like $\sqrt
  m$ for integer $m$ representing the number of vertical nodes in the
  disturbance.

\item BL15 also point out that where no-flow boundary conditions are
  imposed in the vertical direction, the number of unstable surface
  modes (with increasingly finer length scales) increases with higher
  radial resolution, possibly suggesting that the fundamental problem
  in the VSI setup itself could be ill-posed -- at least with respect
  to these surface modes. 

\item As the wavelength of the radially propagating traveling
  wave becomes larger, similarly the growth rate increases in an
  unbounded way.
\end{enumerate}

As with regards to the third deficiency, both numerical simulations of
NGU13 and Stoll \& Kley (2014) indicate that there exists a radial
scale of maximum linear growth yet neither of the analyses of the
asymptotically reduced equations examined by NGU13 nor BL15 admit such
a trend.  Is it possible that the reason for this is due to the
breakdown in the validity of the reduced equations which hinges on the
assumption of radial geostrophy in the dynamics, or might this be a
problem with the assumption of radial traveling waves?  These are
reviewed in more detail in Section~\ref{background}.

In Section~\ref{mode_kinetic_energies}, we argue that the first two of
the above listed troubling features poses no serious deficiency in
either the reduced set of equations or upon the robustness/validity
of the VSI itself. This is because both the surface modes and the
other high nodal modes (i.e. high $m$) carry very little of the total
vertical kinetic energy of the system.  As with regards to the third
issue, we consider this pathology to be a shortcoming of assuming a
radial traveling wave-like solution and not to be a deficiency of the
reduced set of equations.  This in turn is related intimately to
incorrectly assuming separable solution forms for a problem which is
inherently inseparable.  We present an improved approximation in
Section~\ref{improved_solution}, wherein we adopt a relatively
tractable non-separable solution form and reanalyze the reduced
equations.  We find that there exist maximally growing disturbances at
some finite radial length scale and that they, in turn, match the
growth rates and fastest growing radial scales reported in NGU13.  In
Section~\ref{discussion} we briefly discuss our findings.

\section{Background}\label{background}
In both NGU13 and BL15, the following asymptotic reduced set of
equations governing the dynamics of the VSI was shown to be
appropriate in describing the linear development of the instability
for axisymmetric disturbances:
\beqa
0 &=&  2\Omega\sub 0 v - \frac{\partial \tilde \Pi}{\partial x}\,,
\label{rad_geostrophy}\\
\frac{\partial v}{\partial \tau} &=& -\sfrac{1}{2}\Omega\sub 0 \tilde u - \frac{\Omega\sub 0}{2} q z w\,,
\label{azimuthal_balance}\\
\frac{\partial w}{\partial \tau} &=&- \frac{\partial \tilde \Pi}{\partial z}\,, 
\label{vertical_balance} \\
0 &=& \frac{\partial u}{\partial x} + \frac{\partial w}{\partial z} - z w\,.
\label{an elastic}
\eeqa 
With $\Omega\sub 0$ the local rotation rate of the disk section at a
distance $R_0$ from the parent star, this above set was obtained
assuming that the spatial and temporal scales of motion relate
according to the following scalings: temporal dynamics are given by
$\order{1/\epsilon\Omega\sub 0}$, radial dynamic scales $x$ are
$\order{\epsilon^2 R_0}$, and the vertical scales $z$ are on the scale
height $H_0 = \order{\epsilon R_0}$ in which the small parameter
$\epsilon \equiv H_0/R_0$ measures the relative thinness of the disk
(which is usually taken to be approximately $0.05$ in most theoretical
studies including the ones cited above).  The scaled radial and
vertical velocities are $u$ and $w$, respectively, while $v$ is the
deviation azimuthal velocity with respect to the background
near-Keplerian flow, and $\tilde\Pi$ is the scaled pressure
perturbation.  These equations model disk inertial modes with very
short radial wavelengths.  The degree of the vertical shear, which
varies with disk height $z$, is controlled by the parameter $q$ (where
no vertical shear is equivalent to $q=0$).  In both NGU13 and Stoll \&
Kley 2014, the value of $q=-1$ is adopted.  Equation
(\ref{rad_geostrophy}) states that the disturbances are largely in
radial geostrophic balance.  Equations
(\ref{azimuthal_balance}-\ref{vertical_balance}) are the azimuthal and
vertical momentum equations while equation (\ref{an elastic}) is the
anelastic equation of state.  See both NGU13 and BL15 for further
details regarding the derivation of this set of reduced equations.
Because the equations have been appropriately non-dimensionalized,
henceforth we set $\Omega\sub 0 = 1$ on all of the Coriolis and vertical shear terms
appearing in equations (\ref{rad_geostrophy}--\ref{azimuthal_balance}).

This simplified model may then be combined into a single PDE for the
scaled pressure perturbation 
\beq 
-\frac{\partial^2}{\partial
  t^2}\frac{\partial^2 \tilde \Pi}{\partial x^2} + \frac{\partial^2
  \tilde \Pi}{\partial z^2} + \left(1 + q\frac{\partial}{\partial
  x}\right)z\frac{\partial \tilde \Pi}{\partial z} = 0.
\eeq
Assuming normal mode solutions $\tilde\Pi = \hat\Pi(x,z)\,e^{-i\omega
  t} + {\rm c.c.}$ turns the above PDE into the simpler one:
\beq
\omega^2 \frac{\partial^2 \tilde \Pi}{\partial x^2} +
\frac{\partial^2 \tilde \Pi}{\partial z^2} + \left(1 +
q\frac{\partial}{\partial x}\right)z\frac{\partial \tilde
  \Pi}{\partial z} = 0.
\label{simpler_PDE}
\eeq
The problem remains then to construct solutions of this system and
determine the eigenvalue $\omega$ determining the temporal response.
Inspection of the above form shows that this system is inseparable
when $q\neq 0$ - which introduces a number of issues with regards to
linear stability analyses which we discuss briefly below.

One might proceed analyzing this system assuming an approximate
separable form.\footnote{ If the calculation domain is on a uniform
  rectangular grid, a sure way to guarantee an unambiguous
  determination of eigenvalues and eigenmodes is to apply a finite
  difference discretisation of equation (\ref{simpler_PDE}).  With
  $N_z$ and $N_x$ discretisation points in the vertical direction and
  radial direction, then the resulting eigenvalue system requires
  inversion of $(N_x N_z) \times (N_x N_z)$ sized (relatively) sparse
  matrices.  High resolution in both the radial and vertical direction
  are desirable and therefore $N_z \approx 150$, $N_x \approx 100$
  which means constructing matrices that are prohibitively large to
  invert or, equally speaking, challenging to reconfigure into known
  sparse matrix formulations. }  For instance, as was done by BL15,
one can consider the relatively tractable traveling-wave ansatz 
\beq
\hat\Pi = Z(z) e^{ikx}\,.
\label{stupid_ansatz}
\eeq
Then equation (\ref{simpler_PDE}) simplifies further to
\beq
-k^2\omega^2 Z
+ \frac{\partial^2 Z}{\partial z^2} + 
\big(1 + iqk\big)z\frac{\partial Z}{\partial z} = 0\,,
\label{vertical_eqn_wave_ansatz}
\eeq
where $Z(z)$ is an, as yet undetermined, vertical structure function.
The above equation, which is explicitly the same form as appearing in
BL15, is in the form of Hermite's equation.  One solution of this
system that admits tractable analytic results is to allow
perturbations to show (at most) algebraic growth as $|z| \rightarrow
\infty$, in which case \beq Z(z) = He_m(z) \eeq is an acceptable
solution where $m$ is a positive index and He$_m$ is the Hermite
polynomial of order $m$.  The index $m$ counts the number of vertical
nodes in the pressure eigenfunction, and so it is referred to as
indicating this quality throughout the rest of this research note. 
When inserted into equation (\ref{simpler_PDE}), we find that this 
solution form is an actual solution provided the following relationship 
holds:
\beq \omega = \frac{\sqrt{m}}{k}\sqrt{1+ikq}.
\label{simple_dispersion}
\eeq
As discussed in BL15, the growth rate associated with this form
increases without bound as $k \rightarrow 0$.  We agree that this is
pathological for a number of reasons, the following two being most
prominent: (i) the approximation of radial geostrophy most likely
breaks down when the horizontal wavelengths become large, and (ii)
because, as we show in the next section, the traveling wave ansatz is
also deeply flawed.  This solution is also problematic because the
growth rates also grow without bound both as the integer $m$ becomes
large and as $k$ becomes small.  Nonetheless, this simplest solution
indicates that the system is ill-posed as higher-order vertical modes
(increased $m$) are included in the analysis.

Note that this solution ansatz cannot recover surface modes for the
obvious reason that no kinematic boundary conditions are either
emplaced or enforced in the vertical.  We also note that since both
the atmosphere density drops off as a Gaussian (i.e. $\rho\sub 0 \sim
\exp{-z^2/2}$) and that eigenmodes have $z^m$ structure, the effect of
adopting a free vertical boundary condition is really to say that
kinetic energies of all modes decay to zero as $z\rightarrow \pm
\infty$.

Another approach is the one taken by NGU13 and BL15 in which equation
(\ref{vertical_eqn_wave_ansatz}) is solved with no-normal flow
boundary conditions at the vertical boundaries at $z=\pm H$ (note
that $H$ refers to the height of the solution domain and $H_0$ 
refers to the disk scale height in this paper), which
amounts to imposing that $\partial_z \tilde\Pi = 0$ there provided
$\omega \neq 0$, which follows from the normal mode form of equation
(\ref{vertical_balance}).  The general solution of
(\ref{vertical_eqn_wave_ansatz}) is given by
\[
Z(z) = A\, {\rm {He}}_\lambda(z) 
     + B_\lambda {\ }_1F_1\left(-\frac{\lambda}{2}, \frac{1}{2},
\frac{z^2(1+iqk)}{2}\right)
\]
where ${\ }_1F_1$ is confluent hypergeometric function of the first
kind and where $\lambda$ is the usual separation constant which is
determined through the application of boundary conditions.  The second
of these special functions do not offer very much in the form of
analytic ease or insight (NGU13) except in the guise of certain
asymptotic limits (BL15) and, as such, it is more convenient to solve
equation (\ref{vertical_eqn_wave_ansatz}) directly numerically.

\begin{figure}
\begin{center}
\leavevmode
\includegraphics[width=8.8cm]{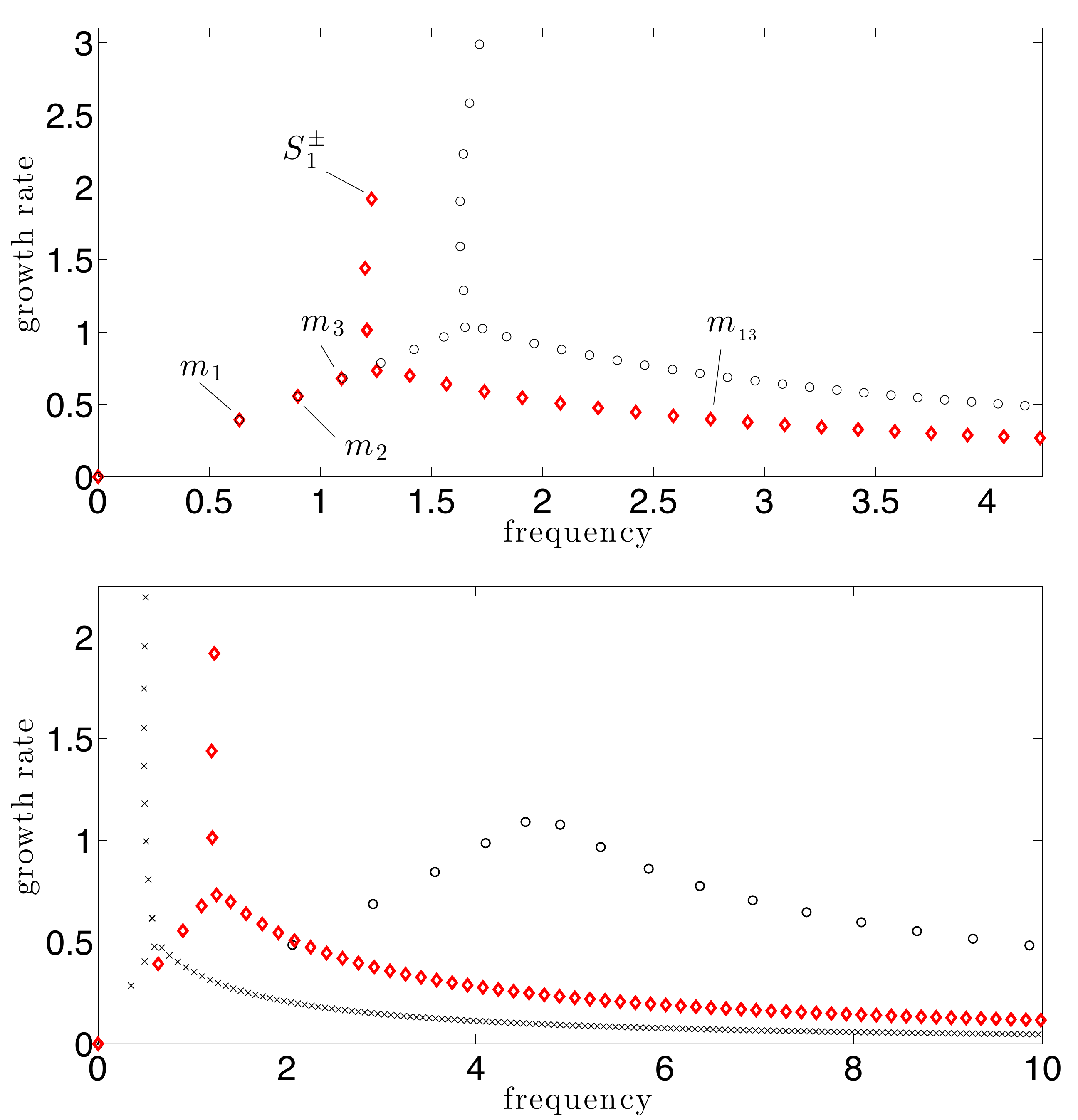}
\end{center}
\caption{The growth rates, Im$(\omega)$, and frequencies,
  Re$(\omega)$, of solutions of equation
  (\ref{vertical_eqn_wave_ansatz}) subject to no-normal flow boundary
  conditions at $z=\pm H$., where $H$ is in units of scale heights.
  {\emph{Top panel:}} Distribution shown for $k=2$ and $H=5$
  (diamonds) and $H=7$ (open circles). 
  As $H$ increases more surface modes become activated and high-order
  body modes have larger growth rates.  In both cases shown, the
  frequency and growth rates of low-order body modes (labeled
  $m_1,m_2,m_3$) remain unchanged.  Note that the surface modes
  generally appear in pairs as indicated by superscript labeling the
  topmost surface mode $S_1^{\pm}$.  This panel confirms the trends
  reported by BL15.  {\emph{Bottom panel:}} Distribution of the
  complex frequencies shown for differing values of $k$ with fixed
  $H=5$: $k=5$ (crosses), $k=2$ (diamonds), $k=0.5$ (open circles).
  The growth rates increase without bound as $k$ is decreased, with
  the same trend found in the problem with no vertical boundaries as
  found in expression (\ref{simple_dispersion}).  Note that as $k$ is
  increased, the number of surface modes increases including the
  maximum growth rates - also confirming the results reported in BL15.
}
\label{travelling_wave_growth_rates}
 \end{figure}

\begin{figure}
\begin{center}
\leavevmode
\includegraphics[width=9.2cm]{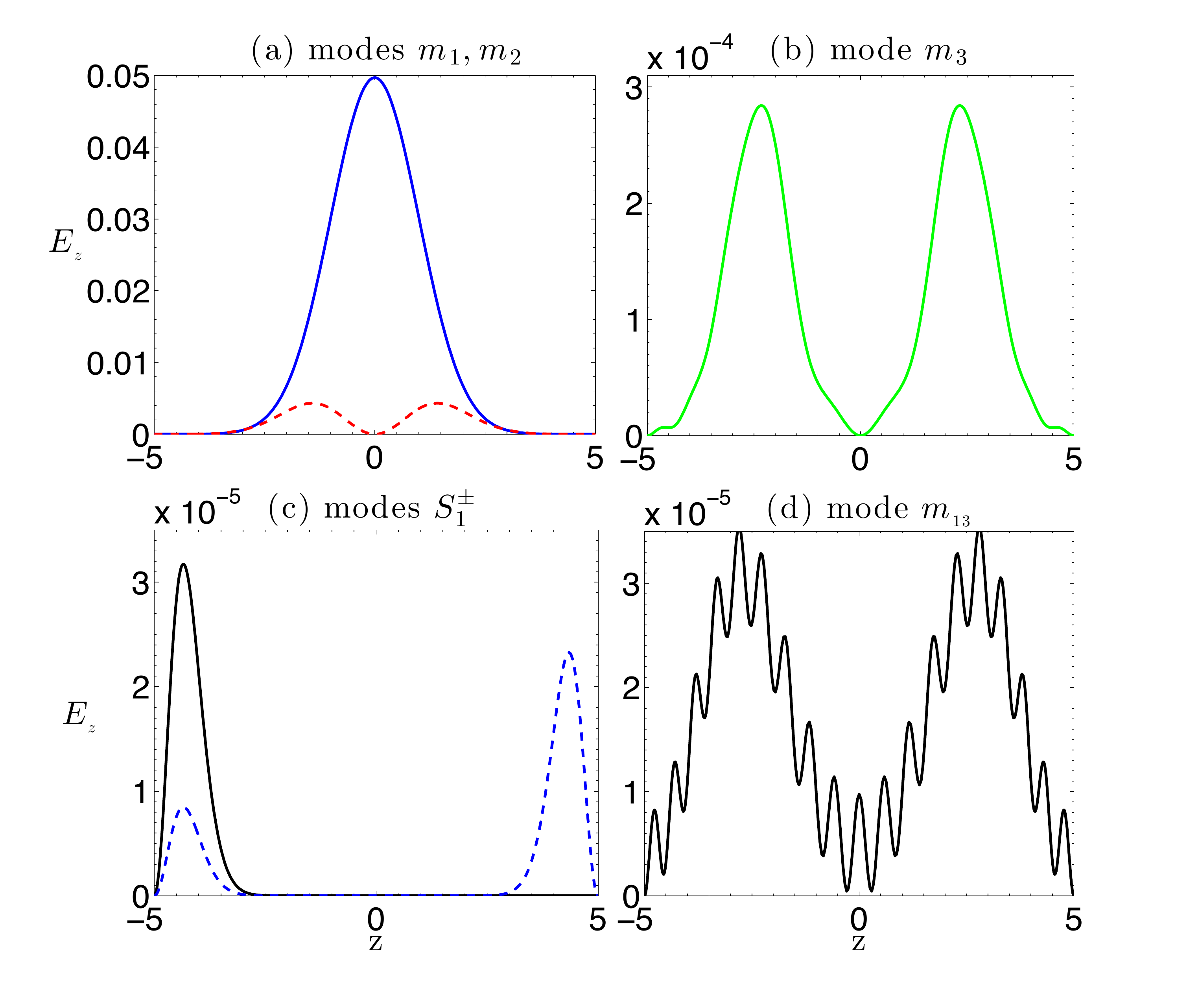}
\end{center}
\caption{Vertical kinetic energy density plots $E_z$ plotted for the
  modes labeled in the top panel of
  Fig. \ref{travelling_wave_growth_rates} corresponding to solutions
  of (\ref{vertical_eqn_wave_ansatz}) with $H=5$ and $k=2$.  Each
  corresponding vertical velocity eigenfunction $w$ is normalized such
  that $\int^5_{-5} |w| dz = 1$. (a) low-order body modes $m_1,m_2$
  (solid and dashed lines respectively), (b) low-order body mode
  $m_3$, (c) fastest growing surface modes $S_1^+,S_1^-$ (solid and
  dashed lines respectively) and (d) a high-order body mode
  $m\sub{13}$.  The lowest order body mode $m_1$ is the fundamental
  corrugation mode and dominates the energy density contained in the
  high-order body modes and the surface modes by at least a factor of
  $10^3$.  The energy density contained in the fundamental breathing
  mode ($m_2$) is a factor 10 less than the mode $m_1$.  }
\label{energy_plottage}
\end{figure}

The numerical eigenvalues determined by this procedure recover the
surface modes as well as the body modes of the VSI.  The low-order
body modes (the fundamental and first overtone breathing and
corrugation modes) are also recovered with eigenvalues consistent with
the numerical results of NGU13.  However this system introduces
apparent pathologies which are depicted in
Figure~\ref{travelling_wave_growth_rates}.  There are generally three
branches of solutions: one associated with low-order body modes with
relatively low frequencies, another branch of body modes with higher
frequencies and a third stem consisting of surface modes.  The high
frequency branch of body modes show decreasing growth rates as the
mode frequency increases while the low-order body modes show
increasing growth rate with increasing frequency.

However, as BL15 demonstrate, when the location of the vertical height
is increased, it is found that (a) the growth rates of the high
frequency branch increase (b) the growth rates of the surface body
modes also increase while (c) the growth rates and frequencies of the
low-order body modes remain unchanged and (d) as the lid of the
atmosphere is raised to $\pm \infty$ the low- and high-order body
modes line up with the frequencies and growth rates found expressed in
equation~(\ref{simple_dispersion}), that is, the response predicted
assuming the ansatz found in equation~(\ref{stupid_ansatz}).  While it
would seem that imposing vertical no-flow boundary conditions lifts
the ill-posedness due to the unbounded growth rates with respect to
$m$ in the body modes arising from the ansatz in
equation~(\ref{stupid_ansatz}), the problem comes back as $H
\rightarrow \infty$ since the growth rates in the high-order body
modes and surface modes correspondingly increase, seemingly without bound.
This is problematic and indicates that the problem may be ill-posed
after all.  

The situation becomes even worse when considering the behavior of the
surface modes, as BL15 indicate: increasing the radial resolution
(larger $k$) proliferates the number of surface modes attached to the
upper and lower no-flow boundaries as the second panel of
Figure~\ref{travelling_wave_growth_rates} clearly illustrates.
Raising the value of the lid also increases the growth rate of the
fastest growing modes also indicating some kind of ill-posedness due
to the surface modes as well.  

Moreover, another serious shortcoming implied by the results of the
linear stability solutions developed in NGU13 and BL15 is that they do
not predict finite, non-zero maximally growing radial wave
disturbances -- something that is however observed in the numerical
experiments of NGU13 and Stoll \& Kley (2014).  This suggests that
assuming wavelike modes in the radial direction is flawed.

\begin{figure*}
\begin{center}
\leavevmode
\includegraphics[width=\textwidth]{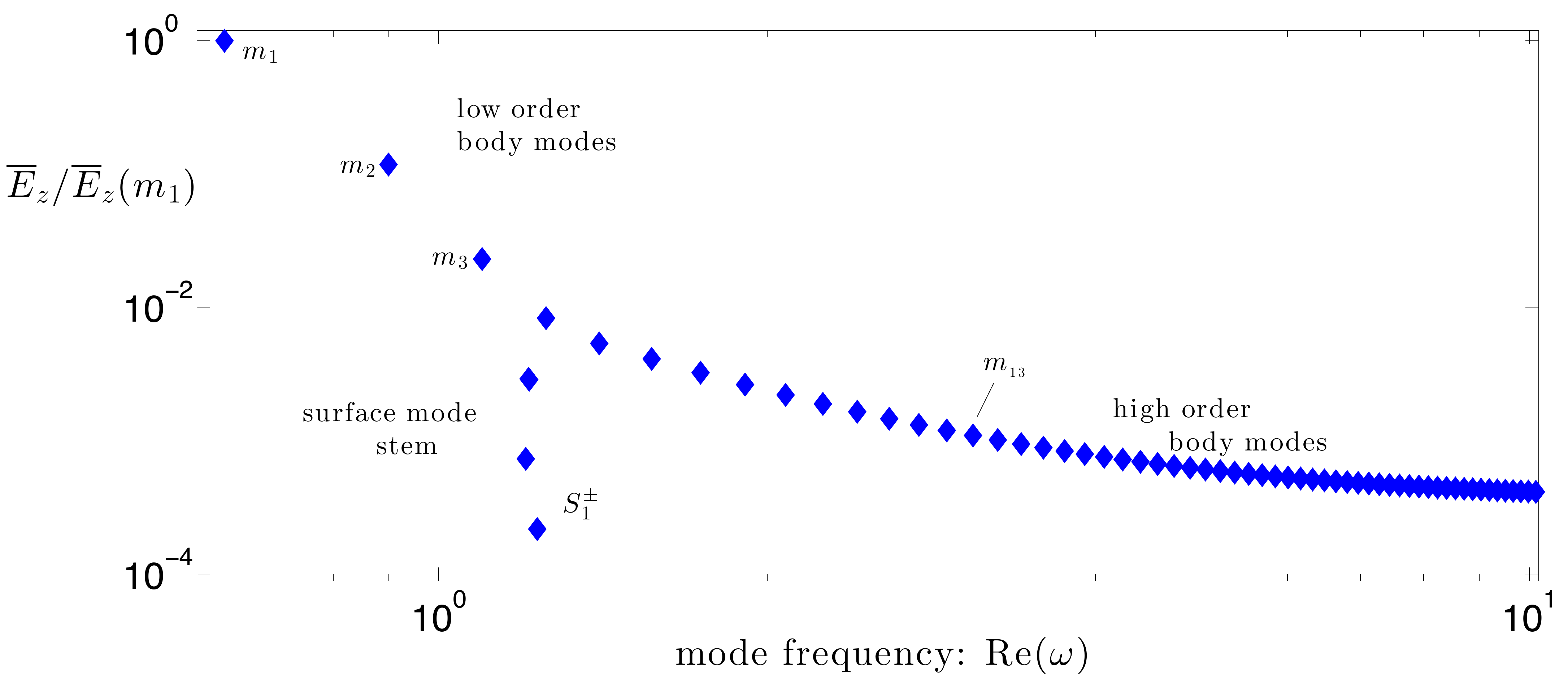}
\end{center}
\caption{Total vertically integrated vertical kinetic energies
  $\overline E_z$ as a function of mode frequencies, Re$(\omega)$,
  corresponding to solutions of (\ref{vertical_eqn_wave_ansatz}) with
  $H=5$ and $k=2$.  $\overline E_z$ is plotted normalized to the
  corresponding vertically integrated energy of the fundamental
  corrugation mode, i.e. $\overline E_z(m_1)$.  Note the relative
  weakness in the power of the surface modes located in the frequency
  window 1.2 and 1.3.  The labeled modes displayed in Figure
  \ref{energy_plottage} are labeled here as well.  }
\label{vertically_integrated_energy_plottageH5k2}
\end{figure*}

\section{Mode kinetic energies}
\label{mode_kinetic_energies}

In spite of these troublesome features, these approximate theoretical
solutions reveal much about the physical nature of the developing
instability - especially with regards to the high frequency body modes
and the proliferating surface modes.  For example, they indicate
something about the relative energy content for each mode. The
low-order body modes carry most of the inertia of the disk
disturbances as their amplitudes are greatest near $z=0$ (NGU13,
BL15).  Since these are also locations where most of the disk mass is
concentrated, the energy contained in these low-order body modes
dominate the corresponding energy contained in the higher-order body
and surface modes.  This has direct consequence with regards to the
interpretation of the VSI, even in the framework of this somewhat incomplete 
analysis.

To quantitatively illustrate this, we show in
Figure~\ref{energy_plottage} a comparison of the relative energy
densities contained in representative modes labeled in the top panel
of Figure~\ref{travelling_wave_growth_rates}, corresponding to
solutions of (\ref{vertical_eqn_wave_ansatz}) with $H=5$ and $k=2$.
From equation (\ref{vertical_balance}) it follows that each eigenmode
$Z(z)$ generates a corresponding vertical velocity eigenfunction $w =
i\omega^{-1} \partial_z Z$.  We normalize each such vertical velocity
eigenfunction so that 
\beq \int_{-H}^H |w| dz =
1\,. \label{vertical_velocity_normalization_procedure} 
\eeq 
Since the reduced equations represent an isothermal atmosphere, the
steady-state density is given by $\rho_0 = \exp\big(-z^2/2\big)$.  We
also recall that this instability is one in which the perturbation
vertical kinetic energy density dominates over the radial and
azimuthal kinetic energy densities (NGU13, Stoll \& Kley, 2014). As
such, we consider the vertical kinetic energy density $E_z$ of each of
the corresponding modes defined by $ E_z(z) \equiv 0.5 \rho_0 |w|^2.
$ The mode labeled `$m_1$' is the fundamental corrugation mode (FCM)
while the mode labeled `$m_2$' is the fundamental breathing mode (FBM)
so that, for example, the expression $E_z(z,m_1)$ corresponds to the
energy density of the FCM, and so on.  We also define for each mode a
total vertically integrated density $\overline E_z \equiv \int^{H}_H
0.5 \rho_0 |w|^2 dz$ (i.e. a surface energy density).  When we refer
to the surface energy density of a particular mode we write, for
example $\overline E_z(m_1)$, to indicate the surface energy density
of the FCM, and so on.

Figure~\ref{energy_plottage} plots the energy densities $\overline
E_z$ for the various modes admitted by the system with parameter
values $k=2, H=5$, wherein each vertical velocity normal mode is
normalized according to
(\ref{vertical_velocity_normalization_procedure}).  We see that the
relative energy density content is greatest with FCM and dominates the
FBM by a factor of 10.  The energy density distribution in the other
higher-order body modes are reduced by at least a factor of 100 compared
to the FCM.  The energy density contained in the two fastest growing
surface modes is diminished by a factor of 1000 compared to the
FCM.

\begin{figure}
\begin{center}
\leavevmode
\includegraphics[width=9.1cm]{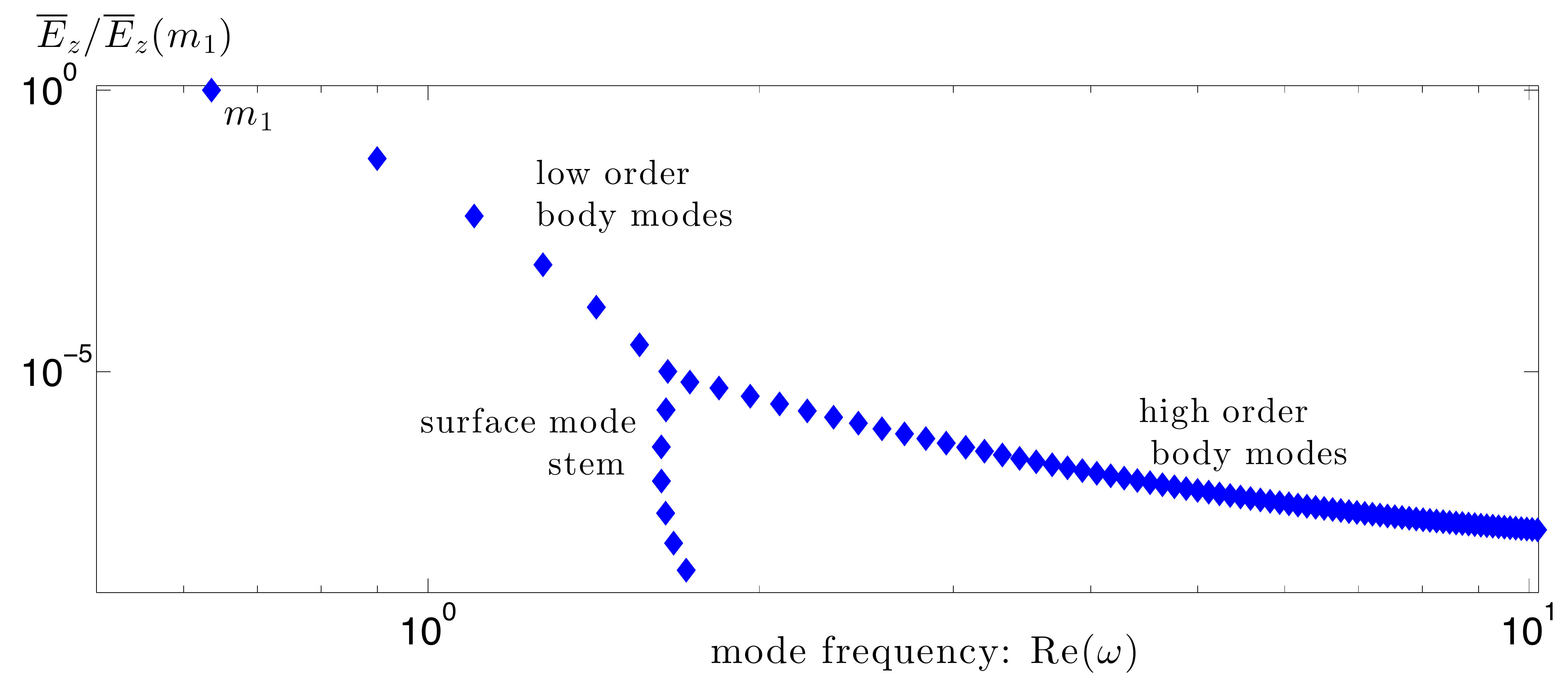}
\end{center}
\caption{Like Figure~\ref{vertically_integrated_energy_plottageH5k2}
  except with $H=7$.  The power contained both in the surface and
  high-order body modes is diminished as the atmosphere lid is set
  further away. Note that the energy in the low order body modes remain unchanged, especially the 
  FCM and FBM.  The only difference is that an increasing number
  of body modes appear and remain stable with increased values of $H$. }
\label{vertically_integrated_energy_plottageH7k2}
\end{figure}

\begin{figure}
\begin{center}
\leavevmode
\includegraphics[width=9.1cm]{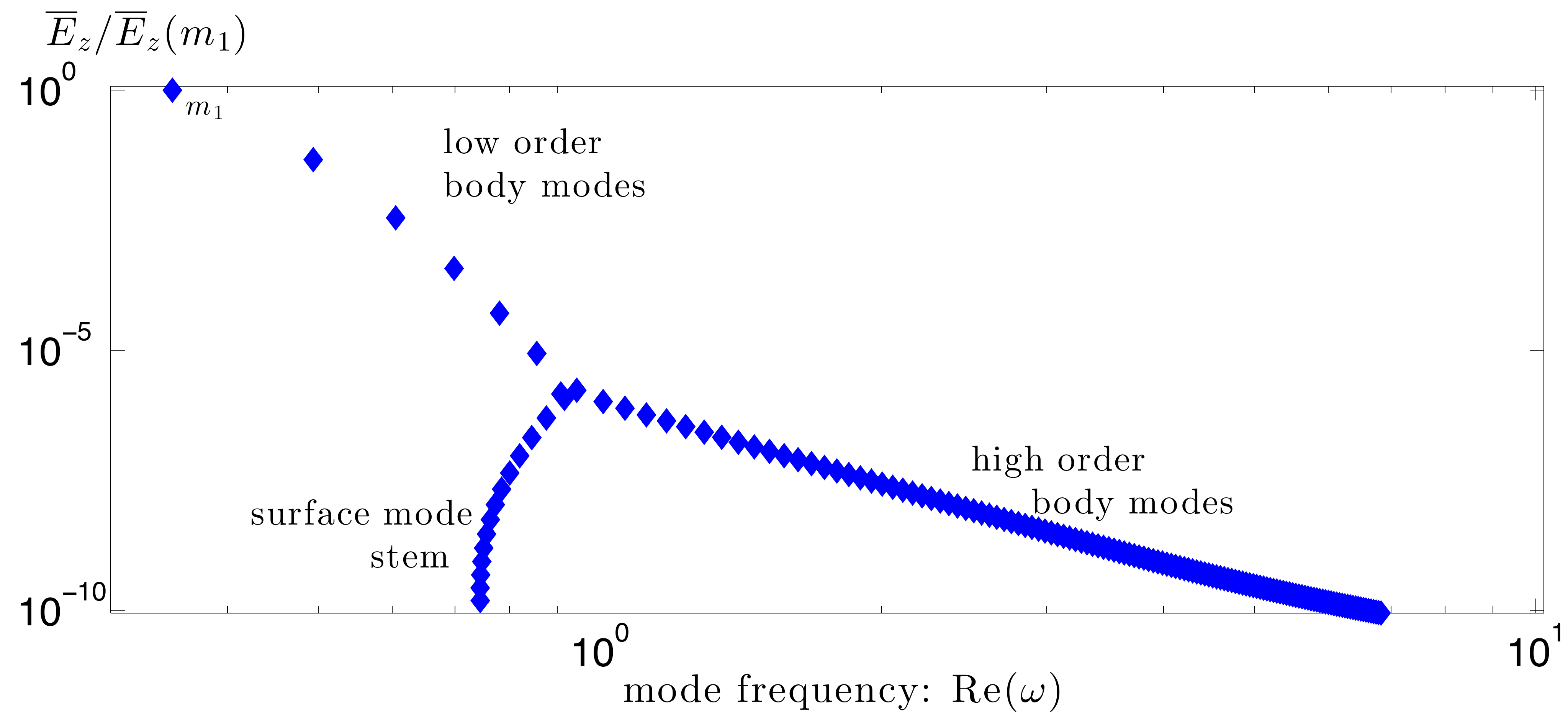}
\end{center}
\caption{Like Figure~\ref{vertically_integrated_energy_plottageH5k2}
except with $H=8$ and $k=5$.  Note the diminished energy carrying
capacity of both the surface modes and the high-order body modes as both
the lid of the atmosphere is made larger and higher radial resolution modes
are considered.  The energy of the low order body modes, especially the FCM, FBM and
the first overtone corrugation mode remain unchanged
compared to their energies for smaller values of $H$ depicted in the two previous figures.
}
\label{vertically_integrated_energy_plottageH8k5}
\end{figure}

A comparison of the total vertically integrated energies in these
various modes emphasizes further the relative unimportance of the
high-order body and surface modes.  Expressing this quantity relative
to the vertically integrated energy density of the FCM ($\overline
E_z(m_1)$), we have for the selected modes: for the FBM, $\overline
E_z(m_2) \approx 0.11 \overline E_z(m_1)$; for the first overtone
corrugation mode, $\overline E_z(m_3) \approx 8.3\times 10^{-3}
\overline E_z(m_1)$; the selected high-order body mode $m_{_{13}}$:
$\overline E_z(m_{_{13}}) \approx 1.3\times 10^{-3} \overline
E_z(m_1)$; and for the two surface body modes, both of which have the
same amount of vertically integrated energy contained within,
$\overline E_z(S_1^{\pm}) \approx 2.2\times 10^{-4} \overline
E_z(m_1)$.  These trends are plotted in
Figure~\ref{vertically_integrated_energy_plottageH5k2} together with
the first 50 vertical eigenmodes, which shows demonstrably that the
energy contained in the modes drops with increased values of $m$.  We
similarly plot the relative vertically integrated kinetic energy
densities for models where $H=7,k=2$
(Figure~\ref{vertically_integrated_energy_plottageH7k2}) and $H=8,k=5$
(Figure~\ref{vertically_integrated_energy_plottageH8k5}) and we see
how increasing the resolution (going to larger $k$) and extending the
atmosphere lid shows how the low-order modes get increasingly
populated (as BL15 point out) and that the $\overline E_z$ energies
contained in the corresponding surface and high-order body modes get
even further diminished. \par
 Most importantly we confirm the trend
reported by BL15 in which the energy in the low order
body modes remains steady as $H$ is increased - this is especially true
for the FCM and FBM but also becomes a characteristic feature of increasing overtones as $H$ is 
taken larger.
In reference to Figure~\ref{vertically_integrated_energy_plottageH5k2}
all the body modes to the left of the triple junction where the surface mode stem branch
meets the low order and high order body modes have energies that are unchanged
as $H$ is increased.  Increasing $H$ however moves the location of the triple junction
toward higher order body modes but the energies of the low order body modes (i.e. those
left of the triple junction) do not change with increased $H$.
\par
The weak relative energy carrying potential of both the high order body modes and the branch of surface
modes comes about because while the high order body modes and surface modes have
have strong power in the vertical velocity field for large values of $|z|$, the corresponding
kinetic energy associated with them gets severely diminished because of the Guassian drop-off associated with the mean
density field $\rho\sub 0$.

\section{An improved approximate solution} 
\label{improved_solution}

While we cannot address all of the concerns enumerated in
Section~\ref{background}, we offer an improved solution ansatz to the
eigenvalue problem posed by equation (\ref{simpler_PDE}).  The
simulations presented in NGU13 and Stoll \& Kley (2014) employ a
numerical set-up in which radial boundaries are enforced.  Such
boundaries introduce effects that alter the growth rates and character
of the low-frequency body modes -- the very modes observed to carry
the instability into the nonlinear regime.  We demonstrate here that
the unbounded growth predicted by assuming wavelike disturbances in
the radial direction is an artifact of assuming radially-traveling
wave solutions and that this pathology is removed by the imposition of
some kind of fixed boundary condition. Furthermore, the imposition of
boundary conditions selects a fastest growing radial wavenumber that
matches the growth rates found in the aforementioned numerical
experiments.

In order to represent the effect of radial boundaries, the following
non-separable ansatz is assumed: \beq \hat\Pi = P_m(z,x) =
\sum_{n=0}^m P\sub{n,m}(x) z^n.
\label{svd_ansatz}
\eeq for $m$ a positive integer and a set of unknown functions of $x$,
$P\sub{n,m}(x)$.\footnote{This ansatz form is non-separable for all
  integer values of $m\ge 2$, and is separable only for the $m=1$
  mode.}  This solution form is one borrowed from singular value
decomposition methods and has been used in other disk studies
(e.g. Lubow \& Pringle 1983).  We note already that the ansatz found
in equation~(\ref{svd_ansatz}) builds into the solutions unbounded
algebraic spatial growth as $|z| \rightarrow \infty$ and will predict
the same kind of unbounded growth in which Im$(\omega) \sim \sqrt m$,
just as the simple solution shown in equation
(\ref{simple_dispersion}). But as we already noted in
Section~\ref{background}, this means these solutions are ones in which
the kinetic energies always decay as $z\rightarrow \pm \infty$.  On
the positive side, these solutions are not burdened by the
introduction of surface modes.

As such, we adopt this solution form and insert it into equation
(\ref{simpler_PDE}).  Separating out like orders in powers of $z$
turns this system into $\order{\frac{m+1}{2}}$ nested ODEs for the
unknown functions $P\sub{n,m}$.  For $n=m$ we have the ``top" equation
\beq \omega^2 \frac{\partial^2 P_{m,m}}{\partial x^2} +
m\left(P_{m,m} + q \frac{\partial P_{m,m}}{\partial x}\right) = 0\,,
\label{top_Pmm_equation}
\eeq
while for $0\le n < m$ we have the remaining ``slaved" equations
\beqa
\omega^2 \frac{\partial^2 P_{n,m}}{\partial x^2}
&+& n\left(P_{n,m} + q \frac{\partial P_{n,m}}{\partial x}\right) = 
\nonumber \\
& & \ \ \ \ \ \ \ \ \ \ \ \ - (n+2)(n+1)P_{n+2,m}\,.
\label{Pnm_equations}
\eeqa
The above system has even and odd symmetries associated with it, so
that there are so-called breathing modes (even $m$) and corrugation
modes (odd $m$).  The fundamental corrugation mode (FCM) corresponds
to $m=1$ while the fundamental breathing mode (FBM) is associated with
$m=2$.

For this particular demonstration, we assume no normal flow boundary
conditions at some inner and outer radial position, i.e. $u = 0$ at
$x=\pm L_x$, where $L_x>0$.  In terms of the variables we use, the
expression of this condition is found by rewriting equation
(\ref{azimuthal_balance}) in terms of the normal mode ansatz \beq
-i\omega \frac{\partial \tilde \Pi}{\partial x} + \frac{qz}{i\omega}
\frac{\partial \tilde \Pi}{\partial z} = 0.  \eeq Given the solution
form (\ref{svd_ansatz}) and since $z^n$ are linearly independent with
respect to one another for integer $n$, each function $P_{n,m}$ must
separately satisfy \beq \frac{\partial P\sub{n,m}}{\partial x} +
\frac{qn}{\omega^2} P\sub{n,m} = 0\quad(\textrm{at}\; x=\pm L_x)\,.
\label{our_bc}
\eeq

The solution method for the full problem is now straightforward: (i)
solve the ``top" equation (\ref{top_Pmm_equation}) subject to boundary
conditions and then (ii) solve the ``slaved" equations
(\ref{Pnm_equations}) subject to the boundary conditions expressed in
equation (\ref{our_bc}) for each $P_{n,m}$ for decreasing values of
$n$ (by 2) until one terminates either at $n=0$ (breathing modes) or
$n=1$ (corrugation modes).  A detailed depiction of the full solution
will be left for a future study.  Of concern to us here, however, is
the fact that the top equation yields the eigenvalue $\omega$.  In
fact, it is straightforward to show that 
\beq \displaystyle P_{m,m} =
\Big(A\sub j\sin k\sub j x + B\sub j\cos k\sub j
x\Big)\exp\left(-\frac{qm}{2\omega^2}x\right), \eeq is a solution to
equation (\ref{top_Pmm_equation}) provided $\omega = \omega(k_j)$
satisfies \beq \frac{\omega^2}{m} = \frac{1 \pm \sqrt{1 - k_j^2
    q^2}}{2k_j^2}, \eeq together with $k = k_j \equiv j\pi/2L_x$,
where $j$ is any integer including zero.  When $j$ is an odd integer
then $A_j = k_j,\ \ B_j = -qm/2\omega^2$ while when $j$ is an even
integer (including zero) $A_j = -qm/2\omega^2,\ \ B_j = -k_j$.  \par
Since all the $k_j$ are real and their multitude are controlled by
$L_x$, we can consider the set of $k_j$ as part of a continuum of real
values given as $k$ and we can analyze these results accordingly.  It
follows that unstable solutions exist only if $|kq| > 1$, and after a
little algebra it implies that the growth/decay rate is given by \beq
{\rm Im}(\omega) = \pm \sqrt{m}\frac{\sqrt{|kq| - 1}}{2k}.
\label{improved_growth_rate}
\eeq
This solution says that there is a wavelength of maximal growth
$k\sub{{\rm max}}$ and corresponding growth rate $\sigma\sub{max}$
which are given by \beq |k\sub{{\rm max}}| = \frac{2}{|q|}, \qquad
\sigma\sub{max}= \sqrt{m}\frac{|q|}{4}.  \eeq Restoring these results
in terms of the physical scalings of the disk, this implies a
maximally growing wavelength $\Lambda_{{\rm max}}$ and growth rate
$\Sigma_{{\rm max}}$ expressed as \beqa & & \Lambda_{{\rm max}} = \pi
\epsilon^2 |q| R_0 = \pi |q| \frac{H_0^2}{R_0},
\label{predicted_fastest_growth_scale} \nonumber \\
& &   \Sigma\sub{max} = \epsilon \sqrt m\frac{|q|}{4}  \Omega\sub 0  = 
\sqrt m\frac{|q|}{4}   \frac{H_0}{R_0} \Omega\sub 0\,,
\label{predicted_fastest_growth}
\eeqa
the latter of which, in terms of local disk orbit times (${\rm orb} =
2\pi/\Omega\sub 0$), is given as $\Sigma\sub{max} = 0.5\epsilon \sqrt
m |q| \pi \ {\rm orb}^{-1}$.

\section{Discussion}
\label{discussion}

The solution developed in Section~\ref{improved_solution} is an
improvement over the previous ones reported in the literature (namely
NGU13 and BL15).  We enumerate below some relevant observations
regarding it.

\begin{enumerate}\itemsep6pt

\item In the numerical experiments reported in NGU13, it was shown
  that in model disks where $\epsilon = 0.05$, and $q=-1$, the growth
  rate of the perturbation kinetic energy during the early phase
  (between 10-25 orbits of the inner disk) of the growing VSI is about
  $0.25 \ {\rm orb}^{-1}$ (see the right panel of their Figure 1.)
  Inspection of the dynamical response during this same phase (see
  their Figure 3) shows primarily a breathing mode character in the
  vertical velocity.  The radial wavelength of the response near the
  left boundary indicates a size of about $0.009 R_0$ (note that this
  corresponds to approximately 17 grid points resolving the fastest
  growing radial mode).  According to the theory developed in the
  previous section, the radial scale and growth rate of the
  fundamental breathing mode (FBM, $m=2$) is given by equation
  (\ref{predicted_fastest_growth}) predicting $\Lambda_{{\rm max}}
  \approx 0.0079 R_0$ together with $\Sigma\sub{max} \approx 0.11
  \ {\rm orb}^{-1}$.  However, the growth rate in the kinetic energy
  is equal to $2\Sigma\sub{max} \approx 0.22 {\rm orb}^{-1}$.  These
  predictions based on this improved approximation compares favorably
  with the results of the numerical simulations.

\item Similarly, for later times in the simulations reported in NGU13,
  the growth rate of the simulations after about 25 orbit times show
  slower growth and the corresponding figures indicate that in this
  latter phase the disk response is primarily that of the FCM.  For
  the FCM with $k$ as given, the theory predicts a growth rate in the
  kinetic energy of about $2\Sigma\sub{max}(m=1) \approx 0.15 {\rm
    orb}^{-1}$ which approximately matches the slowing seen in the
  kinetic energy growth rates displayed in Figure 1 of NGU13.


\item The analysis developed using the solution ansatz in equation
  (\ref{svd_ansatz}) only captures the essence of the low-order body
  modes and cannot say anything about the surface modes
  simply because no vertical boundary conditions are applied.
  However, given our reflections in
  Section~\ref{mode_kinetic_energies} regarding the kinetic energy
  density content of these modes, the surface modes are likely 
  ephemeral having no
  significant dynamical effect upon the development of the VSI in the bulk
  interior where most of the disk inertia is contained.

\item Despite the improved theoretical construction embodied in the
  ansatz of equation (\ref{svd_ansatz}), especially with regards to
  the correct behavior predicted with respect to a fastest growing
  radial mode, the theory still indicates unbounded growth as the
  vertical node number $m$ increases.  Is this indicative of a
  profound flaw in the ansatz or might it be a real effect?  Given our
  reflections upon the lack of energy contained in high-order body
  modes and surface modes (Section~\ref{mode_kinetic_energies}), it
  may be that the theoretical predictions are actually valid and that
  despite the unbounded growth predicted for increasing $m$, the main
  instability and turbulent development in nonlinear calculations is
  driven primarily by the fundamental breathing and corrugation modes.
  \emph{The concomitant fast growing high-order body modes and surface modes
  likely have little effect upon the overall development of the VSI
  primarily because they contain so little energy by comparison to the
  fundamental modes.}

\item Related to the previous point, unless by conspiracy power is
  initially placed only in high-order modes or in the strongly
  localized surface modes, we believe that the VSI develops robustly
  independent of them - even if artificial boundary conditions are
  emplaced on the upper and lower parts of a numerically modeled disk.
  We agree with BL15 that adding a bit of artificial viscosity, or
  possibly a sponge (an artifice commonly used in atmosphere GCMs),
  should either erase or strongly diminish these modes from a
  numerical calculation.  Recall that in both numerical experiments of
  NGU13 and Stoll \& Kley (2014), the velocity fields are shown as
  time snapshots as the instability develops and they show a strong
  initial development in the velocity field near the boundaries due
  primarily to the fast growth of the surface modes.  Yet, the kinetic
  energy densities contained in those disturbances high up in the
  atmosphere are diminished by a factor of $e^{-12}$ due to the
  vanishingly small densities up there.  It is hard to imagine that
  the instability ensuing within the bulk of the disk is dependent
  upon these surface modes and we view them as inconsequential as far
  as the long term development of the VSI is concerned, including its
  aggregate turbulent transport.  This can be verified by new
  numerical experiments in which, for instance, one may replace the
  upper hard boundary by a sponge or something similar.

\item The results of the previous section shows that enforcing
  boundary conditions at the inner and outer radial positions controls
  the growth of the VSI and selects for a fastest growing radial
  structure.  As such, we see the unbounded growth rates resulting
  from the radial traveling wave ansatz (detailed in
  Section~\ref{background}) to be due to the ansatz itself being
  incorrect and less to be because of a breakdown in the validity of
  the assumptions undergirding the equation set
  (\ref{rad_geostrophy})--(\ref{an elastic}), namely, the
  approximation of radial geostrophy of
  equation~(\ref{rad_geostrophy}).
\end{enumerate}
This last point naturally begets the following question: what are the
appropriate boundary conditions to use in such disk models?  How
robust is the VSI to differing radial boundary conditions and how do
these attest to the true conditions of protoplanetary disk dead zones?

\end{document}